\documentclass[manuscript,referee]{aastex}
\usepackage{amsmath}
\usepackage{graphicx}
\shorttitle{Baryonic self similarity.}
\shortauthors{Alard, C.}
\begin{document}
\title{The baryonic self similarity of dark matter.}
\author{Alard, C.}
\affil{Institut d'Astrophysique de Paris, 98bis boulevard Arago, 75014 Paris, France}
\email{alard@iap.fr}
\begin{abstract}
The cosmological simulations indicates that the dark matter haloes have specific self similar properties.  
However the halo similarity is affected by the baryonic feedback.
%
% At the epoch of galaxy formation, the momentum driven winds
%re-shape the dark matter central density core by softening the slope of the density, with a scale
%length imposed by baryonic process.
%
By using the momentum driven winds as a model to represent the baryon feedback, an equilibrium
condition is derived which directly implies the emergence of a new type of similarity.
The new self similar solution has constant acceleration at a reference radius for both dark matter and baryons.
This model receives strong support from the observations of galaxies.
The new self similar properties implies that the total 
acceleration at larger distances is scale free, the transition between
the dark matter and baryons dominated regime occurs at a constant acceleration, and the maximum amplitude of the velocity 
curve at larger distances is proportional
to $M^{\frac{1}{4}}$. These results demonstrate that this self similar model is consistent with
the basics of MOND phenomenology. In agreement with the observations the coincidence
between the self similar model and MOND breaks at the scale of clusters of galaxies. Some numerical
experiments show that the behavior of the density near the origin is closely approximated by a Einasto profile.
\end{abstract}
\section{Introduction.}
The numerical simulations of structure formation in the cosmological context has been providing
us with a large wealth of interesting information about the structure of cold dark matter (CDM)
haloes. Among the results from the numerical simulations two remarkable facts stands out, an universal
dark matter profile (\citet{Navarro1997}, \citet{Navarro2010}, and the occurrence of a power
law behavior 
for the pseudo phase space density, \citet{Taylor}, \citet{Ludlow}.
As an example Ludlow etal (2010) demonstrated that the
residual to the fit of a power law to the pseudo phase space
density $Q(r)=\frac{\rho}{\sigma^3}$ is typically about 10 to 20 \% 
and does not exceed 30 \%. This behavior is observed in a range of about 2 decades. Similar results were also derived by
\citet{Navarro2010} using Aquarius data.
 The power law regime for $Q(r)$ has a well defined outer boundary,
while the inner regime is more difficult to probe due to the intrinsic difficulty of reconstructing the complex
evolution of the phase space density when the system experiences a large number of multi-dimensional folds in phase space.
%
%
 %For a general review of the universal properties of DM haloes see \citet{Navarro2010}}
%
The agreement between the power law
exponent and the prediction from the Bertschinger self similar solution (\citet{Bertschinger}) suggested that the CDM haloes 
had self similar properties. 
However the Bertschinger solution is a purely radial solution which does not corresponds to the haloes
obtained in numerical simulations. The reason for the correct prediction of $Q(r)$ exponent is due to the fact
that Bertschinger's solution belongs to a family of self similar solutions with the same specific values of the constants. The actual
solution belongs to the same family but with completely different density and velocity dispersion, only the pseudo
phase space density and other related quantities are the same in this family of solutions (\citet{Alard}).
Thus Bertschinger's solution should be seen in historical perspective as the first clue to suggest the self similar
nature of the real solution.
A puzzling unsolved problem was the reason for the observed power law behavior of the pseudo phase space density, and 
the non power law behavior of other quantities. 
\citet{Alard} demonstrated that dynamically cold self
similar solutions in a quasi equilibrium situation have  pseudo phase space density with power law behavior.
This result derive from the fact
that the smoothed probability distribution $P(f)$ of $f$ remains self similar (Alard 2013). A direct prediction is that the higher
order moments constructed using $P(f)$ should also be power laws with predictable exponents. These predicted power laws are effectively
consistent with the measurements obtained using data from numerical simulations (Alard 2013). 
Other quantities like the smoothed density
do not have a self similar expectation, and thus are not power laws near equilibrium.
 At this point it is important to note that these results obtained with
 an intrinsically cold dark matter
model are incompatible with self similar models developed in the fluid limit (see for instance \citet{Subramanian}
or \citet{Lapi}). The cold model is not
analog to a continuous model, this fundamental difference is related to the different behavior
of $Q(r)$ and other related quantities, and non self similar smoothed quantities like $\rho(r)$ or $\sigma(r)$. 
Thus $Q(r)$ and higher order moments of the probability distribution of $f$ are fundamental and specific in the cold non fluid approach only.
As a consequence the consistency between self similarity 
and the near equilibrium phase space density corresponding to the collapse of dynamically cold initial fluctuation
is well established. 
This description of pure CDM halos has to be  modified to take into account the baryon population
co-existing with the DM in real galaxies. The baryonic feedback modifies the mass distribution and as a result
influences the DM halo through changes in the gravitational potential.  
This mechanism leads to a much more complex picture. A good illustration of this picture
is given by the recent observations of the rotation curves of galaxies, and in particular from
the THINGS survey (\citet{ThingS})
%
%When baryons and in particular the baryonic feedback is considered the former
%picture is changed. 
%
%
  The high quality rotation curves from \citet{Blok2008}, \citet{Oh2011} show
 that the central density core slope is lower than the expectation from the NFW profile. 
 These strong and indisputable discrepancies between the observations and the expectation from the CDM model 
 have always been a serious problem for the cold dark matter approach. Two different type of solution
 have been proposed to solve this issue, first CDM is not valid and must be replaced by another model, like for
 instance MOND \citet{Milgrom1983}, \citet{Milgrom1986}. Interestingly \citet{Gentile2011} show that the THINGS survey
 rotation curves are consistent with MOND. The second solution is to consider that the baryonic feedback is
 sufficient to produce the softening of the DM density cusp, and this is the solution that will be considered here . The baryonic feedback has a particularly 
 strong effect on the central region of the DM halo by
 flattening the initial density cusp. Using hydro dynamical simulations \citet{OhB2011} show that
 the baryonic feedback model is consistent with the THINGS survey data. 
Since the baryons are coupled to the DM though gravitational interaction, the destruction of the central
DM cusp is produced through rapid fluctuations of the gravitational potential due to the ejection
of gas and dust. This cusp smoothing mechanism has been investigated in details by, \citet{Navarro1996}, \citet{Read}, 
\citet{Pontzen2012}, \citet{Teyssier2013} who demonstrated that this mechanism is efficient and offers a possibility
to reconcile CDM and the observations. The process operates through a cycle of gas ejection and re-accretion, however
it is interesting to note that the re-accretion of the gas does not produce a compensation of the effects on the DM core \citet{Pontzen2012}.
%
% This model works by producing outflow=loss of baryons, which by gravitational coupling modifies the DM halo, Ref's.
% The main source for the outflow are SN and AGN Ref's. **Alard 2011**
%
%%
See \citet{Pontzen2014} for a general review
 on the baryonic feedback model.
% Here the problem of CDM inconsistencies needs to be described, things survey, which by the way support MOND.
%
%In this paper we will adopt the second solution. As a consequence the initial DM cusp is smoothed by the baryonic
%feedback and transformed in a flatter core with a typical scale length imposed by the
%baryonic feedback. 
Note that the observations and the numerical simulations both suggests that central density core is not completely flat.
The problem of the asymptotic slope at the origin of the central core has been studied by,
\citet{Oh2011}, \citet{Pontzen2013},\citet{DiCintio2013}.
%
%
%Since the scale length of the dark matter self similar solution is time dependent, the scale length
%of the baryonic distribution must be also time dependent. However we consider a near equilibrium
%situation where the temporal variation of the baryonic scale radius is small compared to the system typical
%time scale, thus the baryonic and dark matter scale length need only to be asymptotically identical.
%It is expected that the variation of the baryonic scale length is due to the slow accretion of
%new baryonic material. A further development of this situation, is that
%as we shall see if the baryons impose another condition on the dark matter solution the consequence could 
%be the induction of a new similarity class.
%
Imposing a baryonic scale length to the dark matter halo indicates that the scaling relations are affected
by the baryons in a way that may not be compatible with the initial similarity class and
could result in the development new type of baryonic induced similarity. 
This model is supported by the results from high resolution hydro-dynamical simulations, where the baryonic
feedback influence the DM halo parameters and in particular the pseudo phase space density, \citet{Butsky}.
%%
%The dark matter self similar
%solution is time dependent and thus for a given galaxy the distance scale is also time dependent. Since we consider
%a near equilibrium situation the distance scale variations are small compared to the system typical time scale. In a baryonic
%induced self similar regime the temporal variations of the scale radius are due to the variations of the baryonic
%scale radius which can be attributed to the accretion of new baryonic material. 
%%
 The idea of an universal similarity class for the rotation curve of galaxies was already 
introduced by \citet{Persic}, \citet{Salucci1997}, \citet{Salucci2007}, \citet{Donato}. As we shall see this new similarity class
has also some relation to the MOND phenomenology and provide an explanation to the scale independent accelerations observed at a typical 
radius for a large number of galaxies (\citet{Gentile}). 
Note that in the continuation we distinguish between 'scale independence' and 'self similarity'. In the forthcoming sections scale independence
will mean independent of the distance scale (basically the size of the galaxy), while self similarity will mean independent on the scale of
all the variables (distance, velocity, time, see \citet{Alard} Section 3).
\section{The baryonic scaling.}
\label{section_1}
 There is an extensive literature on the baryonic feedback model already existing.
 As an example (\citet{Navarro1996}, \citet{Gnedin}, \citet{Read},
 \citet{Ragone},  \citet{Ogiya}) developed  various models and investigations to document
  the effect of the baryonic feedback on the central parts of the DM halos. These works
  use the effects of supernovae and AGN to drive the baryonic feedback. The effect of
  supernovae tends to be dominant for low mass galaxies ( see \citet{Governato2012} for more details), however there are several mechanisms
  involved in the supernovae feedback. The supernovae outflow may be driven trough "bursty energy transfer" \citet{Governato2012}, or though
  momentum driven wind \citet{Murray}, \citet{Oppenheimer}, \citet{Davé2007}, \citet{Oppenheimer2008}. The model presented here will consider 
  the momentum driven winds as the source of baryonic feedback. Interestingly, \citet{Oppenheimer} found that among 12 others models tested
  the momentum driven outflow model adjusted on local starburst data was the only one to reproduce the inter galactic medium enrichment
  data.
In the momentum driven wind model the
radiation from young stars impinges on dust in the outflow,
which then couples to the gas and propels matter out of the
galaxy.
  The associated outflow of baryonic matter 
 affects the kinematics of the dark halo, with a dominant effect in the central area. This baryonic
 feedback leads to the suppression of the central density cusp and its replacement with a 
 nearly constant  or lower density slope central core.
 As a consequence the scale length of the dark matter halo $r_{DM}$ is imposed by the baryonic processes, and should
 be closely related to the baryonic distribution size $r_B$. 
The foundations of this model are directly derived from the observational
analysis of the rotation curves of galaxies. A correlation between the gas content and the structure
of the gravitational potential is observed \citet{Alard2011}. This analysis is reinforced by the results
of \citet{Lelli2013} and \citet{Lelli2014} who find similar results with the additional finding that
the structure of the potential is also related to the star formation activity. These observational
results show that baryonic process related to stellar formation are responsible for the modification
of the gravitational potential. However the detailed sequence of the events leading to these observational
correlation is not clearly defined. A possibility is that most of the DM halo re-shaping occur at an
early epoch when the star formation is very active, or that a number of particularly strong starburst
has a major infleunce. However,
the observational result from \citet{Kauffmann2014} suggest that the total
amount of energy released by the stellar formation is the real and fundamental parameter.
It is important to note that the detailed sequence
of the events do not really matter as long as the global process remain self similar.
%We consider that the relevant baryonic process are those occurring at the epoch of galaxy formation when the baryonic
%feedback is strong enough to affect the baryonic distribution.
%
% SEE McQuinn 2010 for an estimation of the SFR in dwarfs: on shorter time scale 4 to 20 times the local group dwarf sfr.
% Murray 2005 attempt to procude LM with SFR typical of observations of normal galaxies.
 %Note that the starburst observed when galaxies are forming corresponds to very high star formation rates (SFR) which are at least
 %10 to 100 times the SFR of normal galaxies. 
 %
%%%%%%%%%%%%%%%%%%%%%%%%%%%%%%%%%%%%%%%%%%%%%%%%%%%%%%%%%%%%%%%%%%%%%%%%%%%%%%%%%%%%%%%%%%%%%%%%%%%%%%%%%%%%%%%%%%%%%%%%%%%%%%%%%%%%%%%%%%%%%%%%%%%%%%%%
%%% Potential problem here: total amount of feedback is more important 
%%%%%%%%%%%%%%%%%%%%%%%%%%%%%%%%%%%%%%%%%%%%%%%%%%%%%%%%%%%%%%%%%%%%%%%%%%%%%%%%%%%%%%%%%%%%%%%%%%%%%%%%%%%%%%%%%%%%%%%%%%%%%%%%%%%%%%%%%%%%%%%%%%%%%%%%
{%\bf
%The impact of the baryonic feedback is maximal when the star formation process is at its maximum, which corresponds to the epoch
%  of galaxy formation.  At later time when star formation is less active, the baryonic
% feedback amounts only to a small fraction of what it was at the time of galaxy formation. As a consequence it is clear that most
% of the re-shaping and transformation of the DM halo occur at the epoch of galaxy formation.
}
 
 %It may happen that starburst
 %occur at later time in galaxy evolution, but the observations of galaxies at low redshifts shows that such events are quite rare.
%
%
Actually the dark matter core size $r_{DM}$ is
constructed by a  dominant baryonic process $P$ which has a typical scale length $r_B$, basically $P=P\left(\frac{r}{r_B}\right)$. 
The width at half maximum $r_P$ of $P$ reads, $r_P=P^{-1}\left(\frac{P(0)}{2}\right) r_B$. 
Thus $r_{DM} = k r_B$, with a constant of proportionality $k$ depending on the specific functional form
for $P$. Note that if we had estimated the typical scale length of the process $P$ not by taking the
width at half maximum, but by taking the width at any other fractional value, it would change the value of $k$ but we would still have, $r_{DM} \propto r_B$.
Interestingly, \citet{Donato2} found that the baryonic and dark matter scale length are nearly proportional. 
Similarly a correlation between the rotation curve and the baryonic distribution is also reported by \citet{Swaters2013}.
%
%thus $r_{DM} \simeq r_B$. 
%
The strength of the baryonic 
 influence on the dark matter kinematics must be compared to the gravitational force due to dark matter.
 If the baryonic force dominates we are in a pure flat core regime, while in the case where the dark matter force dominates we return
 to the NFW profile  (Navarro, Frenk, \& White, 1997). The intermediate regime when the dark matter force $F_{DM}$ and baryonic force $F_B$ are of the
 same order defines the typical size of the baryonic core induced regime. The force is proportional to the acceleration
 thus the dark matter core size $r_{DM}$ corresponds to the following condition:
\begin{equation}
 a_{DM}\left( r_{DM} \right)=a_B\left( r_{DM} \right)
 \label{acc_eq}
\end{equation}
Assuming a Burkert profile for the dark matter (\citet{Burkert}), Gentile {\it etal} (2009) found that
at a Burkert profile scale length $r_0$ the respective dark matter and baryonic acceleration are:
$3.2^{+1.8}_{-1.2} \ 10^{-9} {\rm cm \ s}^{-2}$ and $5.7^{+3.8}_{-2.8} \ 10^{-10}  {\rm cm \ s}^{-2}$.
The radius corresponding to a half of the central
value for the Burkert profile is $r_{DM} \simeq 0.55 \ r_0$. Assuming that most of the baryon mass is inside
$r_{DM}$, the baryons in the range $r_{DM}<r<r_0$ behave like a point mass, the acceleration
at $r_0$ must be re-normalized by the scale factor $\left[\frac{r_{DM}}{r_0} \right]^2$ to obtain the acceleration at
$r_{DM}$. On the other hand the acceleration due to the Burkert acceleration decrease by about $13\%$. These
corrections implies that $a_{DM}\left( r_{DM} \right) \simeq 2.8^{+1.6}_{-1.0} \ 10^{-9} {\rm \ cm \ s}^{-2}$ and
$a_B\left( r_{DM} \right) \simeq 1.9^{+1.3}_{-0.9} \ 10^{-9} {\rm \ cm \ s}^{-2}$. These values of the acceleration 
are consistent with Eq. \ref{acc_eq} within the errors bars. This result shows that the scale 
length of the dark matter halo core radius is consistent with the strength of the baryonic influence. 
The radial distance scale of dark matter is imposed by the baryons, which in itself is not sufficient
to impose another similarity class to the dark matter halo. However as we shall see an equilibrium
condition for the gas also implies a constraint on the total force at the distance scale, which is
only compatible with a specific similarity class. 
%of the distance scale (basically the size of the galaxy), while self similar means an independence on the scale of all the variables (distance, velocity, time, see Alard 2013 Section 3).
%
% Due to the self-similar properties of the infall the dark matter halo inherits the same self 
% similar properties. 
%
% However it is interesting to point that if the Baryonic perturbation possess
% specific universal scaling properties these properties may be transmitted to the dark halo through
% the feedback mechanism. 
%
%
\subsection{Critical condition for optically thick gas.}  
\label{section_2}
 We consider the effect of the baryonic feedback on a galaxy composed of stellar populations gas and dark matter. 
%The general
%structure of  dark matter and of the baryonic distribution are constructed at the epoch of the formation of the galaxy. We will
%assume that the dark matter core and the associated scaling properties results from the 
%
%effect of the momentum driven outflows occurring at  this epoch. 
The momentum driven winds offer an efficient mechanism to drive winds other large distances. 
A potential problem in this model is that the gas and dust are ejected from the galaxy and sent to the intergalactic medium.
However to have an effect on the DM halo there must be a cycle where matter is ejected and fall back on the galaxy. But contrary to conventional
wisdom \citet{Oppenheimer2008} show that a cycle occur in the momentum driven wind model, the ejected material is far more likely to fall
back rather than stay in the inter galactic medium, with a typical fall back time of 1 Gyr.
%
% The gas is supposed to be optically thick
%
Note that the critical opacity (optically thick limit) is reached very quickly in a
galaxy due to the production of dust by supernovae. \citet{Murray}
shows that the critical opacity is reached in only $10^6$ years for a
major starburst. As a consequence we will work in the optically thick limit where
the equivalent starburst luminosity $L_B$ of the stellar
 population is entirely absorbed by the gas.
%
%the equivalent starburst luminosity $L_B$ of the stellar
%The starburst luminosity includes the equivalent contribution
% to the transfer of momentum of the 
% of supernovae, the young massive stars and the AGNs. 
%
The mass distribution $M(r)$ is assumed to be spherically
 symmetric. Murray, Quataert, \& Thompson (2005) proposed that such systems are prone to reach a critical luminosity
 $L_B=L_M$. If $L_B>L_M$ (which in general is expected) the acceleration of the gas layer is positive
 and the system loses mass which in turns decrease the star-burst activity. This mechanism operates
 until the star-burst luminosity reach the critical limit $L_M$.
 The luminosity $L_M$ corresponds to an equilibrium between the momentum deposition of the radiation
 in the gas and the gravitational force applied to the gas. 
%
%
% We consider a population of stars located in an optically thick gas layer. 
%The stars have an equivalent star-burst luminosity $L_B$, the gas layer Mass is $M_G$, 
%and the total mass within the shell radius $r$ is $M$. The mass distribution is 
%assumed to be spherically symmetric. As explained in details by Murray, Quataert, \& Todd (2005)
%this system is prone to reach an equilibrium condition with a corresponding luminosity $L_B=L_M$.
%The critical luminosity $L_M$ is defined by the following equilibrium condition (Eq. (14) 
%in Murray, Quataert, \& Todd (2005)): 
%
\subsubsection{Gas distributed in a ring.}
Murray, Quataert, \& Thompson (2005) studied this equilibrium
 for a gas component distributed in a shell. In this case the gas shell diameter is 
supposed to be close to the typical
 size of the galaxy $r_B$, which is proportional to the dark matter core size $r_{DM}$. Considering a total mass of gas $M_G$ 
 the equation corresponding to the equilibrium condition reads:
\begin{equation}
\frac{G M M_G}{r_B^2}  =  \frac{L_M}{c}
\label{gas_eq_ring}
\end{equation}
The starburst luminosity is associated with young stars, and the kinematics of these stars is not much
different from the kinematics of the  cold molecular gas, thus the mass loss affecting this population
of stars will be proportional to the mass loss of the  cold molecular gas. Since we expect that the rate of
new stars will be also proportional to the cold molecular gas mass, the total number of 
stars in the starburst population will be proportional to the cold molecular gas mass.
High quality observations by \citet{Tacconi},  demonstrate that the fraction of cold molecular gas was much higher at the epoch
of galaxy formation than what it is today. On average the fraction of molecular gas is at about 40 \% when
galaxies are forming, which is quite uncommon at the present time. The variability of the molecular gas fraction is also
smaller at the epoch of galaxy formation, which justifies the approximation that the total gas mass and the cold
molecular gas mass are proportional. 
The starburst population goes like the cold molecular gas and that the cold molecular gas mass is approximately proportional
 to the gas mass in forming galaxies.
Thus we consider that the amount of stars in the starburst population will be
approximately proportional to the total gas mass, which corresponds to
the following relation:
\begin{equation}
 L_M \propto M_G
 \label{ML_eq}
\end{equation}
By combining Eqs (\ref{gas_eq_ring}) and (\ref{ML_eq}) we obtain the following equation:
\begin{equation}
\frac{G M}{r_B^2}  =  \rm{Constant}
\label{gas_eq_ring2}
\end{equation}
%
%Note that the assumption made in Eq. \ref{ML_eq} is a simplifying assumption, the proportionality constant in this equation
%may depend slightly on the galaxy type or galaxy mass. Such effects would result also in a slight dependence of the constant
%in Eq. \ref{gas_eq_ring2} on the galaxy type. A trend corresponding to this effect has been documented in the 
%observations by \citet{DelPop}. As a first order approximation
%we will ignore these effects but will keep in mind that corrections due to galaxy type may have to be introduced. However the proportionality factor
%would remain constant during the formation of a given galaxy, basically the galaxy size may somewhat change with time, but
%the gravity field generated by the mass distribution at a typical length scale remains constant.
%
%Hopefully the problem of clarifying the dependence of the star formation
%{\bf
%process as a function of galaxy type should be clarified by the development of new models, which 
%depends strongly on our understanding of starburst activity in galaxies, see \citet{Kaufmann2014}, and
%\citet{Ben} for a general review.}
%
%Hopefully the problem of clarifying the dependence of the star formation
%process as a function of galaxy type should be clarified by the development of new models, however it is important to note that none of the currently
%available models at the moment is a good description of the starburst occurring the epoch of galaxy formation (see for instance \citet{Ben} for a general review).
%
 \subsubsection{General gas distribution.}
\label{Sec3}
In the general case, assuming a density distribution $\rho_G$ of the gas, the equilibrium
at $r=r_B$ condition reads:
\begin{equation}
\int_0^{r_B} \rho_G(r) M(r) dr \propto  \frac{L_M}{c}
\label{gas_eq_gen}
\end{equation}
Applying again Eq. \ref{ML_eq} we obtain:
\begin{equation}
\int_0^{r_B} \rho_G(r) M(r) dr \propto M_G
\label{gas_eq_gen}
\end{equation}
In case of a stable equilibrium,  the total force must be null at the point of equilibrium
but also its first derivative. If we assume that an equilibrium is effectively realized at a typical
baryonic scale $r_B$, we have:
\begin{equation}
\frac{G M}{r_B^2} = {\rm Constant} 
\label{gas_eq_gen2}
\end{equation}
\subsubsection{Consequences of the baryonic critical condition for the self similar solution.}
Eq. (\ref{gas_eq_gen2}) implies that the mass scales like the square of the radius of the distribution which is
not consistent with the Bertschinger self similarity class. The nature of the baryonic scaling implies that
$r_B \propto r_{DM}$, adopting $r_{DM}=\lambda r_B$, using Eq. (\ref{gas_eq_gen2}) and defining a scale free acceleration,
$a(r)=a_2\left(\frac{r}{r_{DM}}\right)$, we have, 
$a(r_B)=\frac{G M}{r_B^2} = a_2\left(\frac{1}{\lambda}\right) = {\rm constant}$. Thus the acceleration is constant at a
fixed point in re-scaled coordinates.
% Using self similar coordinates, $a=\tilde a\left(\frac{r}{r_B}\right)$, thus $\tilde a(\eta)=constant$.
 The self similar regime associated with this constant acceleration corresponds to a specific scaling of the distance and velocity variables. The development of a new
self similar regime induced by conditions developing near the center of the distribution remind us of the of the situation observed for the 
Binney conjecture (\citet{Binney})
. In the two dimensional phase space the Binney conjecture states that for a large
variety of initial conditions the system converges to a power law with an exponent equal to $-\frac{1}{2}$. It was demonstrated by Alard (2013) that 
the power law is induced by a singularity developing at the center of the system. 
Note that since the scale length of the dark matter self similar solution is time dependent, the scale length
of the baryonic distribution must be also time dependent. However we consider a near equilibrium
situation where the temporal variation of the baryonic scale radius is small compared to the system typical
time scale, thus the baryonic and dark matter scale length need only to be asymptotically identical.
It is expected that the variation of the baryonic scale length is due to the slow accretion of
new baryonic material. 
%
% Let's consider a late stage
%of the dark matter halo formation, the growth is slow enough to be almost adiabatic and the regime is asymptotically self similar.
%In this regime Eq. \ref{gas_eq_gen2} implies a specific scaling of the distance and velocity variables. Let first redefine the
%basic features of the analytic self similar solutions of the Vlasov-Poisson system (Alard 2012).
%
%
\subsection{Self similar solutions of the Vlasov-Poisson system.}
 Before we relate the baryonic scaling obtained in the previous section to a given similarity, 
let first remind some of the results obtained in Alard (2013) on the self similar solutions of the Vlasov-Poisson system.
Given a phase space density $f({\bf x},{\bf v},t)$ the general solution in six dimensional phase space reads:
\begin{equation}
 \left\lbrace
 \begin{aligned}
   f({\bf x},{\bf v},t)  & = t^{\alpha_0} g \left (\frac{r}{t^{\alpha_1}},\frac{{\bf v}}{t^{\alpha_2}} \right ) & \\
   \alpha_0  & = -2-3 \alpha_2 \ \ ; \ \ \alpha_1  & = 1+ \alpha_2  &
 \end{aligned} \right.
 \label{self_eq}
 \end{equation}
\newline
Eq. (\ref{self_eq}) implies that the density $\rho$ has the following time scaling:
$$
\rho(r) = \int g(r,{\bf v},t) d^3 v = t^{-2} \rho_2 \left( \frac{r}{t^{\alpha_1}}\right)
$$
Consequently the time scaling of the acceleration reads:
\begin{equation}
 a(r) \propto \frac{M}{r^2} = \frac{1}{r^2} \int \rho(r) r^2 dr = t^{-2+\alpha_1} a_2 \left( \frac{r}{t^{\alpha_1}}\right)
\label{a_eq}
\end{equation}
 The self similar growth of a given dark matter halo with the constraint from Eq. (\ref{self_eq}) implies that the acceleration
at a given re-scaled coordinate remains constant (see Sec. \ref{Sec3}). Thus Eq. \ref{a_eq} should depend only on the re-scaled
coordinate and not on time. Considering the growth of an individual dark matter halo with typical
scale $r_{DM}(t)$, Eq. (\ref{self_eq}) implies that the acceleration at $r_{DM}(t)$ is constant. Assuming a slow
adiabatic process the time dependence of the re-scaling factor in Eq. (\ref{a_eq}) can be linearized. The same
linearization can be applied to $r_{DM}(t)$ which implies that the two expressions become compatible, and that
$r_{DM}(t)$ can be identified to the scaling factor $t^{\alpha_1}$. Considering this identification,
  Eq. (\ref{a_eq}) coupled with Eq. (\ref{gas_eq_gen2}) implies that $\alpha_1=2$, and $\alpha_2=1$. Thus the new
similarity class corresponds to $\alpha_2=1$, this must be compared to the initial Bertschinger similarity class, 
where the similarity was imposed by the nature of the cosmological infall and corresponds to, $\alpha_2=-\frac{1}{9}$. 
\subsubsection{Correspondence between time and distance scales.}
%
%
%The self similar growth of the dark matter halo with the constraint from Eq. (\ref{self_eq}) implies that the acceleration
%at a given re-scaled coordinate remain constant. Since the acceleration in re-scaled coordinates must not depend on time, 
%Eq. (\ref{a_eq}) requires that $\alpha_1=2$ and $\alpha_2=1$. Note that we assumed a self similar solution of the type
%defined in Eq. (\ref{self_eq}), which implies that the time dependence is in the form of power law scaling factors. %Since
%the dark matter typical scale is similar to the baryonic scale $r_B$, it follows that $r_B$ should scale like a power
%law of time. However, it is important to point that we are considering a slowly evolving regime, and that the time
%dependence can be linearized  in both case, for the power law time dependence of the self similar solution (1) 
%and for the baryon induced time evolution of $r_B$ (2), which result in similar time dependencies for (1) and (2). 
%A consequence of the
%halo self similarity class is that the acceleration generated by the dark matter halo $a_{DM}$ does not depend on the scale of the halo. 
The halo velocity and distance scales are related to a free parameter $t_0$ in the definition of time for the self similar solution. 
Basically if $f({\bf x},{\bf v},t)$ is a solution then $f({\bf x},{\bf v},\frac{t}{t_0})$ is also a solution. By introducing $t_0$ the 
scaling of $r$ and $v$ are transformed to respectively , $\left(\frac{t}{t_0}\right)^{\alpha_1}$ and  $\left(\frac{t}{t_0}\right)^{\alpha_2}$. 
The corresponding re-scalings of $r$ and $v$ are $x_0=t_0^{-\alpha_1}$ and $v_0=t_0^{-\alpha_2}$. For the final
stage of the evolution of an individual halo $a_{DM}$ does not depend on time, which also implies that $a_{DM}$  does not depend on 
$t_0$,  and thus does not depend on $x_0$ or $v_0$. These results illustrates the correspondence between time and scale independence. 
Equation (\ref{gas_eq_gen2}) implies that the total acceleration
at $r_B$ is scale free and since $a_{DM}$ is scale free at all positions, the baryonic acceleration at $r_B$, $a_{0B}=a_B(r_{B})$ is scale free or time independent.
The baryonic acceleration at larger distances, out of
the baryonic distribution is properly approximated with the acceleration due to a single massive point, thus,
\begin{equation} 
 a_B \simeq a_{0B} \left( \frac{r_B}{r} \right)^2 \ \ \ r>r_B \ \ \ {\rm with} \ \ a_{0B}=a_B\left(r_B\right)
\label{a0Beq}
\end{equation}
Eq. (\ref{a0Beq}) indicates that at larger distances $a_B$ is only a function of scale free variables and as a consequence is scale free. 
Since $a_{DM}$ is also scale free the total acceleration at larger 
distances ($r>r_B$) is scale free. The fact that the both the baryonic and dark matter accelerations are independent
of scale is supported by the observations. Gentile (2009) found that at a specific scale length $r_0$ ($r_0 \simeq 1.8 \ r_{DM}$),
the baryonic and dark matter acceleration are constant. 
\section{Connection to MOND.}
\citet{Milgrom1983}, and \citet{Beke}, developed an empirical modification of Newtonian dynamics in order
to reproduce the rotation curves of galaxies without the need of including a dark unseen component. The remarkable success
of this approach in reproducing the data (see for instance \citet{MilgromSanders}, \citet{Milgrom2001}, \citet{Milgrom1995}) is particularly compelling since the modelisation is based on parameters reconstructed
directly from the distribution of visible matter. Mond assumes that a transition from the Newtonian regime occurs at an 
acceleration $a_0 \simeq 10^{-10} {\rm cm \ s^{-2}}$ and that below this acceleration we observe an evolution to 
the deep MOND regime which represents the very low acceleration limit. Between the Newtonian and deep MOND regime an empirical
interpolation function is assumed. There are various models for the interpolation function with the obvious consequence that this intermediate
regime is not a very well defined feature of MOND. The essential feature are clearly the acceleration scale at which the transition
occurs and the properties of the deep MOND regime at very low accelerations. These two features derived from MOND are clearly
related to universal properties of galaxies and have to be reproduced by any theory aiming to represent the mass distribution
in galaxies. As we will see the baryonic induced self similar model is consistent with these MOND features. Let's now review MOND general
properties and confront them with the self similar model.
For 
spherically symmetric systems the new equation reads:
\begin{equation}
 \mu \left( \frac{a}{a_0} \right) a =a_B
 \label{milgrom_eq}
\end{equation}
\subsection{Scale free behavior of MOND.}
 \citet{Milgrom1986} already noticed that a similarity relation existed in the MOND approach and that Eq. \ref{milgrom_eq} can be written in a
 dimensionless form (see Milgrom 1986, Eq. 5). At larger distances $r>r_B$ it is straightforward to re-write Eq \ref{milgrom_eq} using
Eq. \ref{a0Beq}:
\begin{equation}
 \mu \left( \frac{a}{a_0} \right) a =a_{0B} \left ( \frac{r_B}{r} \right)^2
\label{milgrom_eq2}
\end{equation}
The constant $a_{0B}$ is independent of scale thus the acceleration in Eq. \ref{milgrom_eq2} is only of a function of scale free variables.
In the self similar model the dark matter acceleration is scale free, the baryonic acceleration (Eq. \ref{a0Beq})
is scale free at larger distances, thus the total acceleration is scale free at larger distances ($r>r_{B}$), which is consistent
with Eq. (\ref{milgrom_eq2}).
\subsection{Rotation curves in MOND and the self similar model.}
\begin{figure}
\includegraphics[width=150mm]{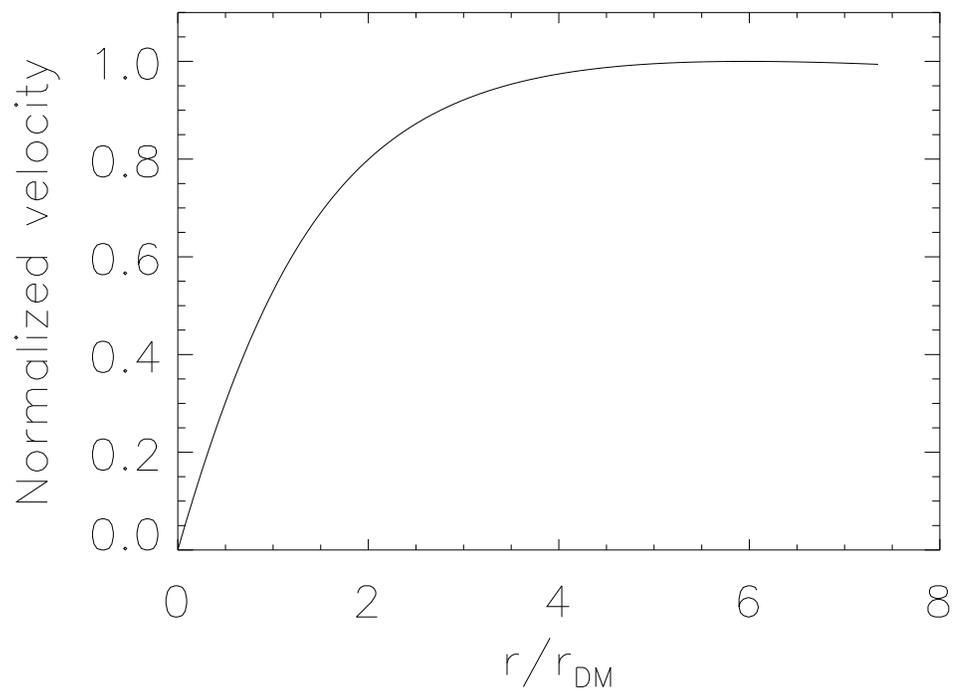}
\caption{The rotation velocity associated with a Burkert profile. The maximum of the velocity curve was
normalized to unity. The radial coordinates is scaled using the the Burkert profile scale length $r_{DM}$.}
\label{Fig0}
\end{figure}

 A general feature of the MOND phenomenology is that at larger distances ($r \gg r_B$, and $a \ll a_0$), the Newtonian force field is a point
mass field which implies that the relation between the velocity at large distances, $v_M$ and the baryon mass $M_B$ is:
\begin{equation}
 v_M^4=a_0 G M_B
\label{mond_vel}
\end{equation}
The velocity at $r \gg r_B$ in the self similar model corresponds to the maximum of the velocity curve $v_M$, which for
a typical dominant dark matter profile, occurs at larger distances (see Fig. \ref{Fig0} for an illustration corresponding to a Burkert profile). We define the position
the position of the maximum of the self similar velocity curve $r_M$, with, $r_M =\eta r_{DM}$, then $\frac{v_M^2}{r_M}=a_M$, is scale free, and  we obtain:
\begin{equation}
v_M^4=\eta^2 \frac{a_M^2}{a_B(r_{DM})} G M_B
\label{self_vel}
\end{equation}
An identification between Eq. \ref{mond_vel} and Eq. \ref{self_vel} indicates that
 $a_0=a_B(r_{DM}) \left(\frac{a_M}{a_B(r_{DM})} \right)^2 \eta^2$. We will adopt 
$a_B(r_{DM}) \simeq a_{DM}(r_{DM}) \simeq 2 \ 10^{-9} \ {\rm cm \ s}^{-2}$ (see Sec. \ref{section_1}). 
Note that $\frac{a_M}{a_B(r_{DM})}=\frac{a_2(\eta)}{a_2(1)}$ is scale free and thus a constant, since the acceleration is scale free at larger
distances. Assuming that an estimation of the constants can be obtained by modeling the mass distribution with a Burkert profile, 
$\frac{a_M}{a_{DM}(r_{DM})} \simeq \frac{1}{2}$ and $\eta \simeq 6$, we obtain, 
$a_0 \simeq 1.8 \ 10^{-8} \ {\rm cm \ s}^{-2}$. Milgrom (2001) estimated that $a_0$ is of the order of $10^{-8} \ {\rm cm \ s}^{-2}$ which is consistent, and shows that the MOND and the self similar model have the same expectation at large distances.  An additional point is that the self similar model predicts 
that at a characteristic scale $r_{DM}$ the acceleration due to the baryons is of the order of the acceleration due to dark matter. The scale free behavior of the
acceleration implies effectively that a the distance scale $r_{DM}$ the acceleration
is constant. The region $r \simeq r_{DM}$ corresponds to the MOND intermediate regime,
where the function $\mu(x)$ is between the Newtonian regime $\mu=1$ and the deep MOND regime, $\mu=x$ . The fact that the transition between the baryon dominated 
regime and the dark matter regime occurs at a fixed acceleration in the self similar model is a clear connection to MOND. To compare the acceleration $\bar a$ at 
which the transition occurs in the two approach we have to consider the equivalent of the situation where $a_{DM}=a_B$ in the MOND approach.
Considering Eq. \ref{milgrom_eq} 
this will corresponds to $\bar a=2 {a_B}$, which in turn implies,
\begin{equation}
\mu\left(\frac{\bar a}{a_0}\right)=\frac{1}{2}
\label{mu_eq}
\end{equation}
Begeman et al. (1991) showed that a sample of high quality rotation curves of galaxies could be fitted using $\mu=\frac{x}{\sqrt{1+x^2}}$ and
$a_0=1.2 \pm 0.27 \ 10^{-8} \ {\rm \ cm \ s}^{-2}$. Using these results the soultion of Eq. \ref{mu_eq} is $\bar a \simeq 0.69 \pm 0.16  \ 10^{-8} \ {\rm \ cm \ s}^{-2}$.
The results
of Sec. \ref{section_1} implies that in the self similar model $\bar a = a_{DM}(r_{DM})+a_B(r_{DM}) \simeq 0.47^{+0.21}_{-0.13} \ 10^{-8} \ {\rm cm \ s}^{-2}$ which is
consistent with the MOND value for $\bar a$ considering the error bars. We compared the large distance low acceleration and intermediate regime  between MOND and
the self similar model. In the remaining domain ($a \gg a_0$) the dynamic is Newtonian and since in galaxies this
regime also occurs at shorter distances from the center ($r \ll r_{DM}$) the baryons dominates and will thus satisfies the Newtonian
limit of MOND (Eq. \ref{milgrom_eq}). As a consequence the asymptotic limits in the MOND and self similar approach are the same, the difference
is only a matter of interpolation between
the low acceleration and Newtonian limits. In MOND the interpolation function itself is not defined
in the theory and is free to vary within some limited constraints. However there is a significant
difference between the self similar model and MOND, the equilibrium equation (\ref{gas_eq_gen2})
applies to a galaxy,
but some fundamental mechanisms are missing to apply it to clusters of galaxies. 
Despite the fact that core formation via a feedback due to AGN has been found to
operate in clusters of galaxies (\citet{Martizzi2013}), the nature of the process
does not include a regulation mechanism that would lead to an equilibrium
condition like Eq. (\ref{gas_eq_gen2}). In galaxies the regulation operates 
by interaction between star formation and the loss of gas. If star formation
is too high the wind are higher that the critical limit, which implies that
gas is removed and as a result slow down star formation (\citet{Murray}). 
Such regulation mechanism does not exists with the AGN feedback model of \citet{Martizzi2013}.
As a consequence this self similar model and 
its associated phenomenology should not be present in
cluster of galaxies, which is a clear difference with MOND. This break of the phenomenology  is in good agreement
with the observations as illustrated with the case of the Bullet cluster (\citet{Clowe2006}, and \citet{Clowe2004} ). 
\section{Density and pseudo-phase space density of dark matter halos.}
\label{Section_5}
It was demonstrated in Alard (2013) that the pseudo-phase space density of self-similar solutions of the Vlasov-Poisson system has a power law behavior.
When this result is coupled with the
Jeans equation an equation for the density can be obtained (see \citet{Dehnen} for a discussion in the case of the Bertschinger similarity class).
This section will now discuss the solution for the density in the case of the baryonic induced similarity class. It is clear that changing
the similarity class has a major influence on the solution for the density, and that the work already conducted for the similarity
class $\alpha_2=-\frac{1}{9}$ needs to be re-conducted for the new similarity class $\alpha_2=1$.
We will assume a spherically symmetric system, thus the spatial coordinates will be reduced to the radial distance modulus $r$.
The Jeans equation reads:
\begin{equation}
 % \frac{1}{\rho(r)} \frac{d}{dr} \left( \rho(r) \sigma(r)^2 \right)+\frac{2 \beta(r) \sigma(r)^2}{r} + \frac{G}{r^2} \int_0^r \rho(u) u^2 du
 \frac{1}{\rho} \frac{d}{dr} \left( \rho \sigma^2 \right)+\frac{2 \beta \sigma^2}{r} + \frac{G}{r^2} \int_0^r \rho u^2 du=0
 \label{jeans_eq}
\end{equation}
The pseudo-phase space density $\frac{\rho}{\sigma^3}$ is a power law with predictable exponent for the self similar solutions
of the Vlasov-Poisson system (Alard 2013). The exponent is a function of the self similar solution constant $\alpha_2$,  
$\frac{\rho}{\sigma^3} \propto \left(\frac{r}{t^{\alpha_1}}\right)^{\gamma}$, and $\gamma=-\frac{2+3\alpha_2}{1+\alpha_2}$.
In the regime of the imposed baryonic self similarity, $\alpha_2=1$, thus $\gamma=-\frac{5}{2}$. It is useful to define the density and anisotropy
parameter as a function of the re-scaled variable, $u=\frac{r}{r_0}$:
\begin{equation}
\rho(r)=\rho_0 \rho_s\left(u \right) \ \ ; \ \ \beta(r)=\beta_s\left(u \right) ; \ \ \sigma(r)= \sigma_0 \sigma_s\left(u \right)  \ \ ; \ \ M(r)= \rho_0 r_0^3 M_s\left(u \right)
\label{rho_sig}
\end{equation}
\newline
The Jeans equation in these variables reads:
 \begin{equation}
  5\,u{\frac {d^{2} M_s}{d{u}^{2}}}  -5\,{\frac {d}{du}}M_s
  +6\, \beta {\frac {d M_s}{du}}
   +3 q_0 \left[\frac{\frac{d M_s}{d u}}{u}\right]^{\frac{4}{3}}=0
 \label{beta_eq}
 \end{equation}
By a applying a derivative to Eq. ~\ref{beta_eq} an equation for the density $\rho$ is obtained.
\newline
\begin{equation}
  \rho_s \left( 15 u \frac{d^2 \rho_s}{du^2}+\rho_s \left(18 \frac{d \beta_s}{du} +\frac{1}{u} \left(48 \beta+40 \right)\right)+\frac{d \rho_s}{du}\left(65+12 \beta \right)\right)-5 u \frac{d \rho_s}{du}^2 +q_0 \rho_s^{\frac{7}{3}} u^{-\frac{2}{3}}=0 
  \label{rho_sig2}
\end{equation}
\newline
With the following definition for the parameter $q_0$:
$$
q_0=\frac{9 G \rho_0 r_0^2}{\sigma_0^2}
$$
Assuming that the halo is virialized at radius $r_0$, we obtain an estimation of the parameter $q_0$.
$$
 \int G M \rho r dr \simeq \int \rho \sigma^2 r^2 dr
$$
With:
$$
\frac{\rho}{\sigma^3} = \frac{\rho_0}{\sigma_0^3} u^{\frac{5}{2}}
$$
The former equation reads:
\begin{equation}
 q_0 \int G M_s \rho_s u du = \int \rho_s^{\frac{5}{3}} u^{\frac{11}{3}} du
 \label{q0_eq}
\end{equation}
Equation \ref{q0_eq} provides a direct estimation of $q_0$. 
\subsection{General solution and asymptotic properties.}
 There are two types of asymptotic regimes to consider, a power law or a constant core.
 \subsubsection{Power law asymptotic behavior.}
 The power law solution for a similar equation was already discussed extensively by Dehnen
 \& McLaughlin (2005). The asymptotic behavior at origin is related to the dominant behavior
 of the left term in Eq. \ref{jeans_eq}, which implies an asymptotic solution of the type
 $\rho \sigma^2 \equiv {\rm constant}$. As a consequence, with $\rho \propto r^{\alpha}$
 and $\frac{\rho}{\sigma^3} \propto r^{\gamma}$ the corresponding asymptotic behavior is
 $\alpha=\frac{2}{5} \gamma$. For the Bertschinger solution, $\alpha_2=-\frac{1}{9}$, $\gamma=-\frac{15}{8}$,
 and $\alpha=-\frac{3}{4}$ which corresponds to the results obtained by Dehnen
 \& McLaughlin (2005). For the solution discussed in this paper $\gamma=-\frac{5}{2}$, and $\alpha=-1$.
 Note that $\alpha=-1$ is the limit for the dominance of the left term in Eq. \ref{jeans_eq}, and that
 as a consequence we have a full solution of the equation, not just for the dominant term.
\subsubsection{Constant core asymptotic behavior.}
\begin{figure}
\includegraphics[width=150mm]{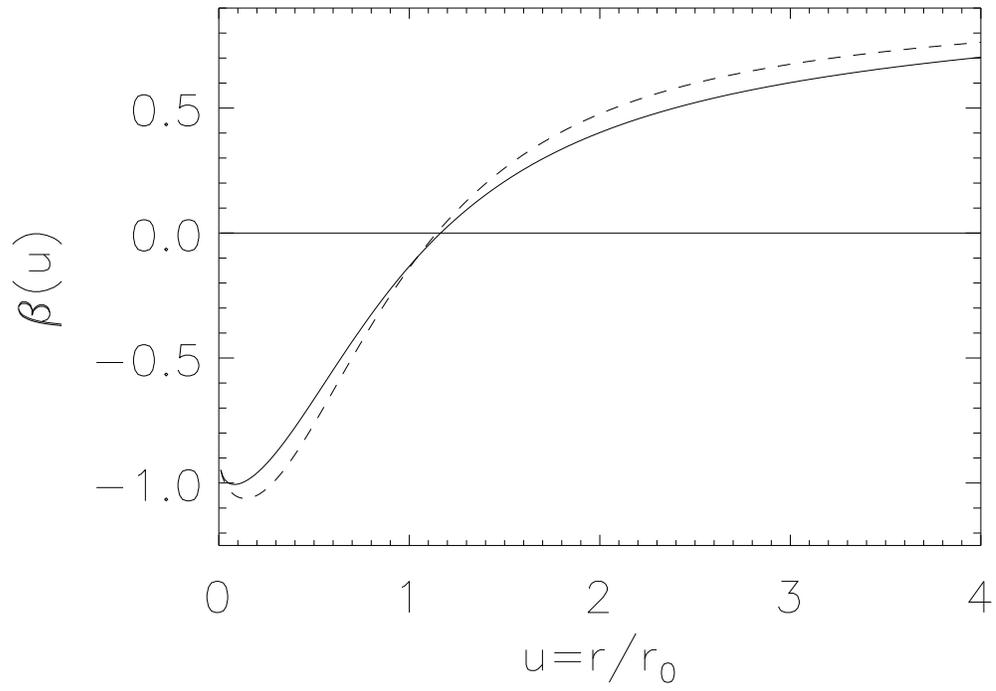}
\caption{The anisotropy parameter $\beta$ for the cored isothermal profile (dashed line) and
for the Burkert profile.}
\label{Fig1}
\end{figure}

 The observations favor models with constant density core, like cored iso-thermal models \citet{Spano} or
 the Burkert profile (\citet{Burkert}). In such case the general solution of Eq's \ref{beta_eq} and \ref{rho_sig2}
 writes:
 \begin{equation}
 \left\lbrace
 \begin{aligned}
  \rho_s & = \sum_{n=0}^{\infty} a_n u^{\frac{n}{3}}\\
  \beta_s & = \sum_{n=0}^{\infty} b_n u^{\frac{n}{3}}  &
  \end{aligned} \right.
 \label{exp_sol1}
 \end{equation}
It is interesting to write explicitly the first terms of the solution series expansion:
\begin{equation}
 b_0=-\frac{5}{6} \ \ \ b_1=\frac{\frac{5}{3} a_1+q_0 a_0^{\frac{4}{3}}}{6 a_0} \ \ \ b_2=\frac{(100 a_2 a_0+7 q_0 a_0^{\frac{4}{3}} a_1-50 a_1^2)}{180 a_0^2}
\label{exp_sol2}
 \end{equation}
A general property of the solutions presented in Eq. \ref{exp_sol2} is that the zeroth order term in the expansion of $\beta$ is constant. Another
point is that in general the next terms in the expansion are of low order in $u$, unless these terms are equal to zero, the asymptotic behavior
of $\beta$ at origin will not correspond to a local minimum of the function. 
An approximate estimation of the functional $\beta$ is obtained by assuming a simple empirical model known for its good consistency with
the observations, like the cored isothermal model:
$$
\rho_s \propto \frac{1}{(1+u^2)^{\frac{3}{2}}}
$$
Or the Burkert profile:
$$
\rho_s \propto \frac{1}{(1+u)(1+u^2)}
$$
The functional $\beta$ is directly estimated by introducing these models of the density in Eq. (\ref{beta_eq}), the results are presented in Fig. \ref{Fig1}.
The variable $q_0$ is estimated using Eq. (\ref{q0_eq}), we find $q_0 \simeq 3.2$ for the cored isothermal density and $q_0 \simeq 3.4$ for the
Burkert profile. Both profiles converge at $\beta=-\frac{5}{6}$ at the origin and cross the zero line near $u=1$, at larger distances increase
slowly and converge to radial orbits at infinity, which is consistent with the cosmological infall. Although these profiles allows us to reproduce
the general features of $\beta$, a generic problem is that in both case the minimum of $\beta$ is not situated at origin. Another way to consider
the problem would be to assume a generic behavior for $\beta$ and estimate $\rho$. The fixed properties of $\beta$ are the value at the origin
and the crossing of the zero line at $u=1$. If we add that the minimum of $\beta$ must be located at the origin, it is straightforward to infer 
a parabolic model for $\beta$.
\begin{equation}
\beta=\frac{5}{6}(u^2-1) \ , \ \ \ 0<u<1
\label{beta_mod}
\end{equation}
By introducing Eq. (\ref{beta_mod}) in Eq. (\ref{rho_sig2}) we obtain an equation for $\rho$. The solution of this differential equation
is obtained numerically using a Runge Kutta method. Finding the solution requires a value of $q_0$, but since $q_0$ is unknown at the initial step,
a first guess is assumed for $q_0$, a solution is found and $q_0$ is estimated. This process is iterated until the guess for $q_0$ and the value
estimated from the numerical solution are the same. We start from $q_0=3$ which corresponds approximately to the values obtained for the cored isothermal
and Burkert profiles. Starting from this initial value the iteration process converges to a value of $q_0=5.59$. 
The numerical solution for $\rho$ corresponding to the simple asymptotic model of $\beta$ at origin described in Eq. (\ref{beta_mod}) is closely
approximated with a Einasto profile (\citet{Einasto}):
\begin{equation}
\rho(r) \propto \exp\left(e_0 r^{\frac{1}{n}}\right) 
\label{Einasto}
\end{equation}
Note that
the nature of the expansion in Eq. (\ref{exp_sol1}) implies that if the solution is consistent with an Einasto profile, then we have necessarily $n=3$.
\begin{figure}
\includegraphics[width=150mm]{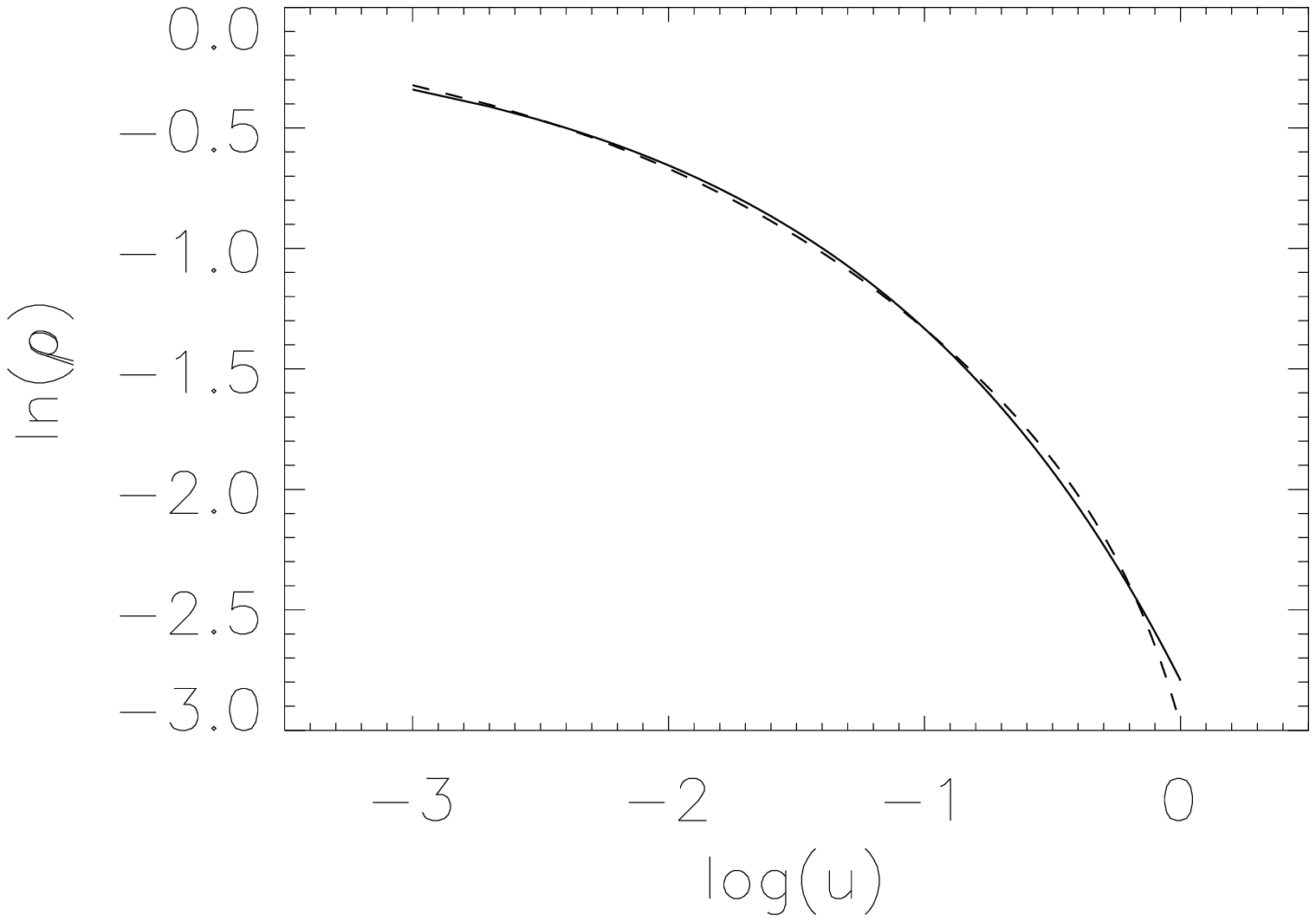}
\caption{The dotted line represents the density obtained by solving Eq. \ref{rho_sig2} for a functional $\beta$ given by Eq. \ref{beta_mod}.
The continuous line is the adjustment of an Einasto profile ($n=3$) to the numerical solution.}
\label{Fig2}
\end{figure}
\section{Synthesis and conclusion.}
 The main concept presented in this article is that the initial dark matter self similarity is affected by the baryon feedback 
 and replaced with a baryonic induced self similarity. It was demonstrated in Sec. (\ref{section_2}) that the baryonic feedback impose two conditions
on the dark matter distribution. These baryonic constraints are not compatible with the initial similarity
class. Provided that the solution remains self similar, the conditions from Sec. (\ref{section_2}) imply the emergence of a
new similarity class for the dark matter halo.
An important point is that this model relies on the assumption that an equilibrium is obtained between the wind pressure and the gravity of the system, leading to a baryonic
induced similarity of the DM halo. However to reach this equilibrium the equivalent luminosity $L$ must be greater than some critical luminosity $L_M$ (\citet{Murray}).
Obviously if the star formation in the galaxy is not sufficient to reach this critical luminosity, no universal acceleration would exists and the associated self similarity
class would not be present. In this case it is not even clear that any self similar properties would emerge from the baryonic feedback. But we must keep in mind
that violating the equilibrium condition would definitely go against the observations, and the universal accelerations for the galaxies observed by \citet{Gentile}.
This would also again go against the general MOND conjecture \citet{Milgrom1983}, \citet{Milgrom1986}, \citet{Milgrom1995}, \citet{Milgrom2001}. Another crucial assumption
is the proportionality between the luminosity and the mass of gas (see \ref{section_2}, Eq. \ref{ML_eq}). An open possibility is that the scale factor between mass and luminosity 
in Eq. \ref{ML_eq} depends on galaxy type. As a result we would still have a baryonic induced self similarity for each galaxy, but the similarity parameter (the constant in Eq.
\ref{gas_eq_ring2}), which is an acceleration would depend on galaxy type. Interestingly \citet{DelPop} found that the acceleration constant estimated by \cite{Gentile} 
is correlated to the mass of the galaxy, which would support the fact that the scaling in Eq. \ref{ML_eq} depends on galaxy type. A direct consequence of this finding
is also to support the fact that the nearly universal relations observed for galaxies are due to internal physics within the galaxies, a category to which the baryonic
feedback obviously belongs. A final and crucial assumption is that the effect of the baryonic feedback is sufficient to induce a new class of similarity
in the DM halo. It is reasonable to consider that the baryonic feedback is sufficient to alter the shape of the DM core and that this process has self similar properties.
But does this means that this baryonic self similarity is transmitted to the whole DM halo ? It is clear that at least self similarity should be transmitted in some
domain with boundaries scaling like the typical of the scale of the baryonic distribution. However, does this means that the baryonic self similarity will be transmitted 
to the very central region, does self similarity breaks at some small fraction of the baryonic scale ? We should also expect that self similarity breaks at some distance in the outer regions. At the moment
the answer to these questions is not clear, but hopefully some new insight should come from the detailed exploration of this type of model using numerical simulations. A possible
observational test of self similarity is provided in Sec. \ref{Section_5}, with the prediction of an Einasto profile with index $n=3$. However the reconstruction of the parameters
of an Einasto profile is especially difficult do to the intrinsic difficulty of subtracting the baryon contribution in the inner region (\citet{Chemin2011}). 
%
%occurring when the galaxy is forming. This assumption is well motivated since on average, the baryonic feedback during galaxy formation is much stronger than at the present epoch.
%It is also clear that the gas fraction in galaxies \citet{Tacconi} is much larger at the epoch of galaxy formation on average than at the present epoch. 
%Thus consuming an amount of gas as large as during the initial starburst would not be likely at later times, and would result in much less
%probable powerful baryonic feedback after the initial episode for most galaxies.
%
%
One important property of this new self similar solution is that the acceleration
generated by the dark matter halo is scale free. When combined with the properties of the baryonic feedback, the scale free 
acceleration of the dark matter implies that the baryon acceleration at one scale radius $r_{DM}$ is independent on $r_{DM}$.
This self similar model put a number of observational facts on galaxies in a coherent framework. First, the universality of the
baryon and dark matter accelerations observed at a scale radius of the dark matter distribution for a large number of galaxies
(Gentile {\it etal} 2009), and second this self similar model is related to the MOND phenomenology of galaxies (Milgrom 1983, and Bekenstein \& Milgrom 1984).
An additional point is that the
density corresponding to this self similar model is expected to form a flat cored distribution in the
central region, and a large variety of profiles has been proposed to fit the observations, cored isothermal, Burkert, or Einasto profiles.
Among these possibilities, the Einasto profile has the best 
compatibility near the origin with the expectation from the self similar, baryonic induced model. 
In these results it is of particular interest
to point out that the in self similar model CDM and MOND become consistent with each other
. The general features observed in the rotation curves of galaxies are properly described in the MOND framework, but we see that it
can be as well represented by dark matter self similar solution.

 The incompatibility of the MOND phenomenology and of the observations in general is a serious problem
for the cold dark matter model (See for instance \citet{Kroupa} for a review). Thus the result that this new cold dark matter model based on self
similarity is consistent with the observed phenomenology is definitely a change in the CDM paradigm. 
A major difference
between the MOND approach and the self similar CDM model is that the self similar model does not apply to clusters
of galaxies, since the equilibrium condition (Eq. \ref{gas_eq_gen2}) does not apply to a cluster. The 
discrepancies between the MOND phenomenology and the observations of the Bullet cluster are thus predicted by the self
similar model.
\acknowledgements
The author would like to thank J.P. Beaulieu and S. Colombi for comments, and the referee for interesting
suggestions. This work has been funded in part by ANR grant ANR-13-MONU-0003.
{}
%
%
%\bibitem[Gilmore etal. (2009)]{Gil} Gilmore, G., Grebel, E., Koch, A., 2009, MNRAS, 397, 1169
%
\end{document}